\documentclass[aps, prx, amsfonts, superscriptaddress, amssymb, amsmath, reprint, showkeys, nofootinbib, raggedbottom]{revtex4-1}
\usepackage[english]{babel}
\usepackage[utf8]{inputenc}
\usepackage{amsthm}
\usepackage{mathtools}
\usepackage{physics}
\usepackage{xcolor}
\usepackage{graphicx}
\usepackage{adjustbox}
\usepackage{placeins}
\usepackage[T1]{fontenc}
\usepackage{lipsum}
\usepackage{csquotes}
\usepackage{float}
\usepackage{afterpage}
\usepackage{booktabs}
\usepackage{bm}
\usepackage[version=4]{mhchem}

\newcommand{\qlineSI}{S1}
\newcommand{\SANSdiffuse}{S2}
\newcommand{\powdersus}{S4}
\newcommand{\chisqvsparams}{S8}
\newcommand{\MCPD}{S9}
\newcommand{\MCdiff}{S10}

\usepackage{ulem}

\usepackage[pdftex, pdftitle={Article}, pdfauthor={Author}]{hyperref} 
\graphicspath{ {Figures/} }

\begin{document}
\title{A Fermi Surface Driven Spiral Spin Liquid}

\author{Paul M. Neves}
\email{pmneves@mit.edu}
\affiliation{Department of Physics, Massachusetts Institute of Technology, Cambridge, MA 02139, USA}
\author{Chi Ian Jess Ip}
\affiliation{Department of Physics, Massachusetts Institute of Technology, Cambridge, MA 02139, USA}
\author{Takashi Kurumaji}
\affiliation{Division of Physics, Mathematics and Astronomy, California Institute of Technology, Pasadena, CA 91125, USA}
\author{Shiang Fang}
\affiliation{Department of Physics, Massachusetts Institute of Technology, Cambridge, MA 02139, USA}
\author{Joseph A. M. Paddison}
\affiliation{Neutron Scattering Division, Oak Ridge National Laboratory, Oak Ridge, TN 37830, USA}
\author{Lisa M. DeBeer-Schmitt}
\affiliation{Large Scale Structures Section, Neutron Scattering Division, Oak Ridge National Laboratory, Oak Ridge, TN 37830, USA}
\author{Daniel G. Mazzone}
\affiliation{PSI Center for Neutron and Muon Sciences, CH-5232 Villigen PSI, Switzerland.}
\author{Jonathan S. White}
\affiliation{PSI Center for Neutron and Muon Sciences, CH-5232 Villigen PSI, Switzerland.}
\author{Joseph G. Checkelsky}
\affiliation{Department of Physics, Massachusetts Institute of Technology, Cambridge, MA 02139, USA}

\date{\today} 

\maketitle
\section{Abstract}

\ce{EuAg4Sb2} is a model material to study the interplay of electronic and spin texture degrees of freedom, exhibiting numerous multi-$q$ magnetic textures coupled with the electronic properties. It is generally understood that some combination of conduction-electron mediated interactions, frustration, and higher order interactions are responsible for complex incommensurate spin textures in centrosymmetric lanthanide materials. Here, we refine an effective model of the magnetic interactions in \ce{EuAg4Sb2} through measurements of diffuse magnetic neutron scattering above the ordering temperature. These diffuse measurements reveal a ring of fluctuating spin modulations that reflects a manifold of nearly degenerate propagation vectors known as a spiral spin liquid (SSL).
We further identify that this approximate U(1) symmetric SSL emerges from magnetic interactions mediated by a quasi-2D hole pocket and exhibits critical scaling of the spatial correlations. Further, Monte Carlo simulations reveal excellent agreement with experiment and provide a comprehensive understanding of the phase diagram. This study emphasizes the connection between the rich spin textures in this material, the electronic structure, and spin liquidity---uncovering new insights into design principles for nano-scale spin texture materials with advantageous intertwined electronic, magnetic, and topological properties, and new mechanisms for generating the physics of spiral spin liquids.

\section{Main}
Spiral spin liquids (SSLs) are correlated paramagnets characterized by hosting a manifold of degenerate propagation vectors in reciprocal space \cite{bergman2007order, yao2021generic}.
Such states are proposed to host novel low-energy fluctuations such as emergent gauge theories \cite{yan2022low}, exhibit unusual entropy driven 'order-by-disorder' phases\cite{bergman2007order}, and may provide insight into the nature of quantum spin liquids \cite{niggemann2019classical}.
Most known SSL materials are insulators such as \ce{MnSc2S4} \cite{gao2017spiral, iqbal2018stability}, \ce{FeCl3} \cite{gao2022spiral}, \ce{LiYbO2} \cite{graham2023experimental}, \ce{GdZnPO} \cite{wan2024spiral}, and \ce{Ca10Cr7O28} \cite{takahashi2025spiral}. These are driven by a fine-tuned balance of $J_1$-$J_3$ exchange interactions which give rise to a manifold of degenerate states. \ce{AgCrSe2} is a notable exception \cite{andriushin2025observation}, but it is also understood in terms of a standard competing $J_1$-$J_3$ model.
All of these known SSLs rely on a fine tuned frustration between near-neighbor exchange energies. Here, we discuss a different conduction-electron mediated SSL formation mechanism relevant to a broad class of technologically relevant conducting materials.

Recently, great attention has been set on lanthanide intermetallics which exhibit a variety of complex magnetic textures due to their potential spintronic applications \cite{kurumaji2019skyrmion, tokura2020magnetic, shimizu_spin_2021}.
In such materials, the conduction electron mediated spin interaction is critical to the stability of these phases \cite{ozawa2016vortex, ozawa2017zero, wang2020skyrmion, hayami2021noncoplanar, nomoto2020formation, bouaziz2022fermi}. In particular, such modulations create a kinetic energy benefit by opening gaps in the electronic bandstructure when the  magnetic propagation vector $\bm{q}$ is matched to twice the Fermi momentum $2k_\text{F}$ \cite{altshuler1995criticalbehaviorofthet}. This bandstructure picture is relevant in the limit where the electronic mean free path exceeds the spin modulation period. In general, this mechanism also can favor multi-$q$ states which more completely gap the Fermi surface \cite{solenov2012chirality, wang2020skyrmion}. The wide variety of multi-$q$ spin textures, or spin moir\'e superlattices  (SMS) \cite{shimizu_spin_2021} makes these materials appealing for spintronic applications due to the possibility to engineer the band dispersion or even generate topological edge states \cite{hamamoto_quantized_2015, zhang2020skyrmion}. This is further exemplified by comparison with Recent work stabilizing exotic phases in 2D heterostructures \cite{andrei_marvels_2021, geim2013van, novoselov20162d} including unconventional superconductivity \cite{cao2018unconventional} and the fractional anomalous quantum Hall effect \cite{cai2023signatures, xu2023observation, lu2024fractional} with Moir\'{e} potentials.

\ce{EuAg4Sb2}, a rhombohedral semimetal hosting triangular lattices of Eu$^{2+}$ $S=7/2,\ L=0$ moments, is an ideal such case where the quasi-2D $\alpha$ pocket yields multiple SMS phases, all with $q\sim2k_F$ (experimentally, $2k_F$ was previously determined to be 0.152$\pm$0.005 \AA$^{-1}$) \cite{kurumaji2025electronic}. Upon cooling, the material enters two different double-$q$ vortex lattice phases starting at $T_N=10.7$ K, and then enters the ground state single-$q$ cycloidal phase \cite{neves2025polarized}. Additional multi-$q$ states can be accessed with the application of in-plane field \cite{green2025robust, neves2024grasp}. As the spin modulations gap the Fermi surface at $q\sim2k_F$, these incommensurate magnetic phases (ICMs) are closely intertwined with the electronic properties of the material \cite{kurumaji2025electronic, neves2025inplane}. The wide variety of phases with highly variable magnetic propagation vectors, all with $q\sim2k_F$, hints at an energy landscape with a remarkable degree of near-degeneracy---a feature we expect to be common amongst related compounds. In this paper we explore this rich landscape in detail, revealing that the conduction-electron mediated interactions give rise to a metallic spiral spin liquid state above $T_N$.

\begin{figure*}[ht]
\includegraphics[width=1.0\textwidth]{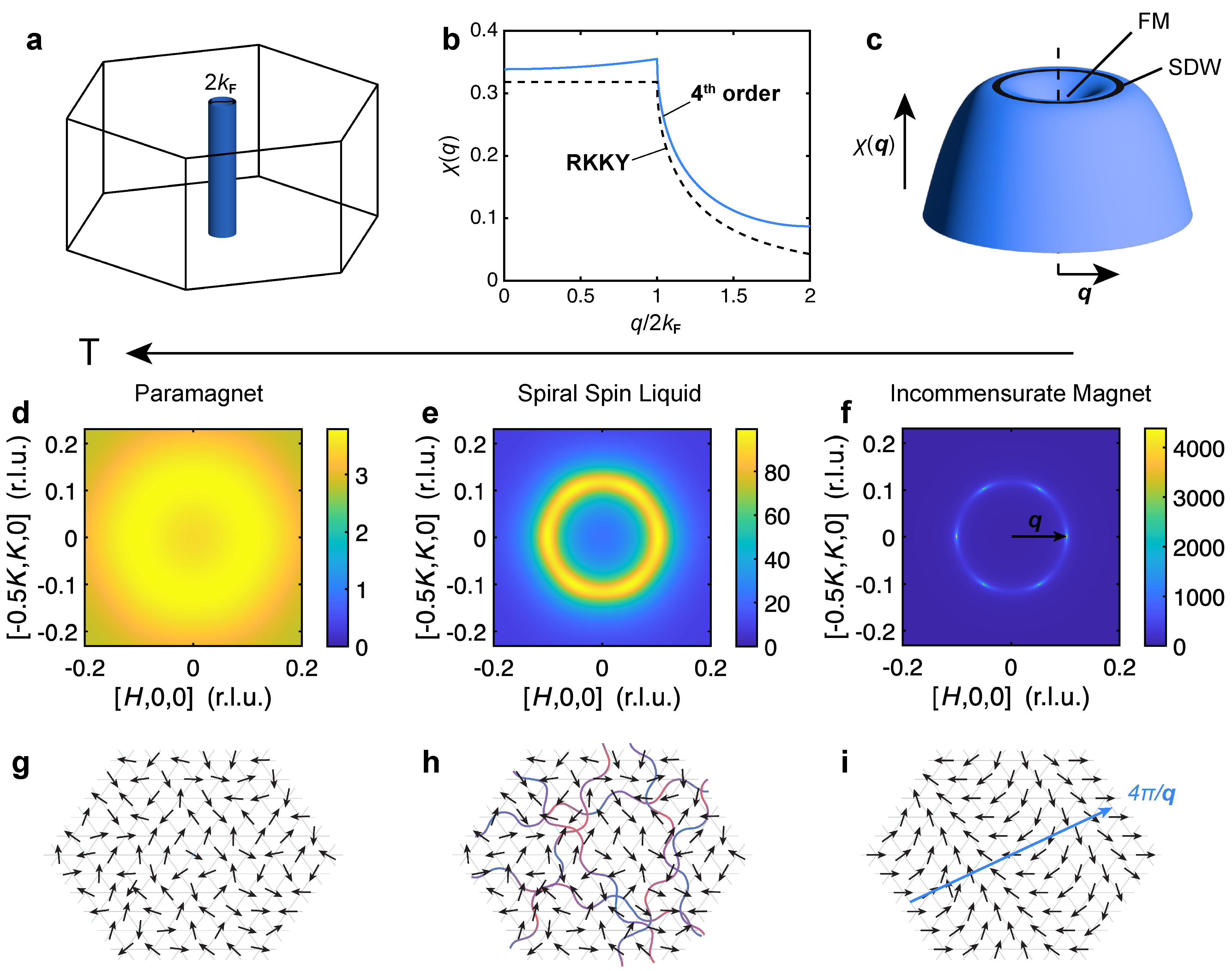}
    \caption{
    \textbf{A Fermi Surface Mediated Spiral Spin Liquid}
    \textbf{a} Schematic 2D Fermi surface in a hexagonal Brillouin zone.
    \textbf{b} RKKY-mediated momentum dependent susceptibility for a parabolic band in one, two, and three dimensions (after \cite{kittel1969indirect}). The general concept that more refined approximations can favor $q=2k_F$ order in 2D is depicted schematically with a dashed black line.
    \textbf{b} Magnetic susceptibility of a 2D electron gas treated to second order (dashed line) and fourth order (solid blue line), after \cite{wang2020skyrmion}.
    \textbf{c} Schematic illustration of a case proximate to the 2D RKKY case where the ferromagnetic ($\bm{q}=0$) spin susceptibility is slightly less favored than a 2D line of $|q|=2k_F$ in momentum space.
    \textbf{d-f} Model momentum-space resolved susceptibility and \textbf{g-i} schematic of 2D SSL at three representative temperatures depicting the high temperature paramagnetic, intermediate temperature spiral spin liquid, and low temperature cycloidal ordered behavior, respectively.
    }
    \label{intro}
\end{figure*}

As it is believed that the small, approximately cylindrical $\alpha$ pocket (see Fig. \ref{intro}a) predominantly mediates the magnetic exchange energy \cite{kurumaji2025electronic}, the indirect exchange interactions are approximately isotropic within the triangle plane, and can be quite long-range.
In other words, opening a gap for any spin modulation with $q\sim2k_F$ in any direction in the plane is approximately the same energy (see Fig. \ref{intro}b,c), which we will see leads to a SSL phase above the ordering temperature (see Fig. \ref{intro}d-i).
This represents a new mechanism for obtaining a SSL phase that does not rely on fine tuning.
Here, by modeling the observed diffuse neutron scattering above the ordering temperature, we will find an effective model which replicates the low energy landscape of this induced interaction and resultant SSL phase.
This model also accurately predicts experimental observations of the ordered state which were not included in the input dataset.
The mediation of a SSL via the cylindrical isotropic Fermi surface in \ce{EuAg4Sb2} can be seen as a new material platform for the controlled generation of such phases, and more broadly a new insight into the formation mechanism of spin moir\'e superlattices with desirable electronic band features.

\section{Evidence for SSL State}

Previous neutron scattering work on \ce{EuAg4Sb2} has considered the magnetic scattering observed in the diffraction peaks associated with the long range order present below $T_N=10.7$ K \cite{kurumaji2025electronic, green2025robust, neves2025polarized, neves2025inplane}. Here, we consider in detail the diffuse intensity that forms before the long-range order is established. While magnetic Bragg peaks represent long-range order, a system with short range spin correlations will still exhibit diffuse scattering intensity in the region of momentum space corresponding to the wavelength of the spin fluctuations present (see supplementary section SII for additional information).

\begin{figure*}[ht]
\includegraphics[width=1.0\textwidth]{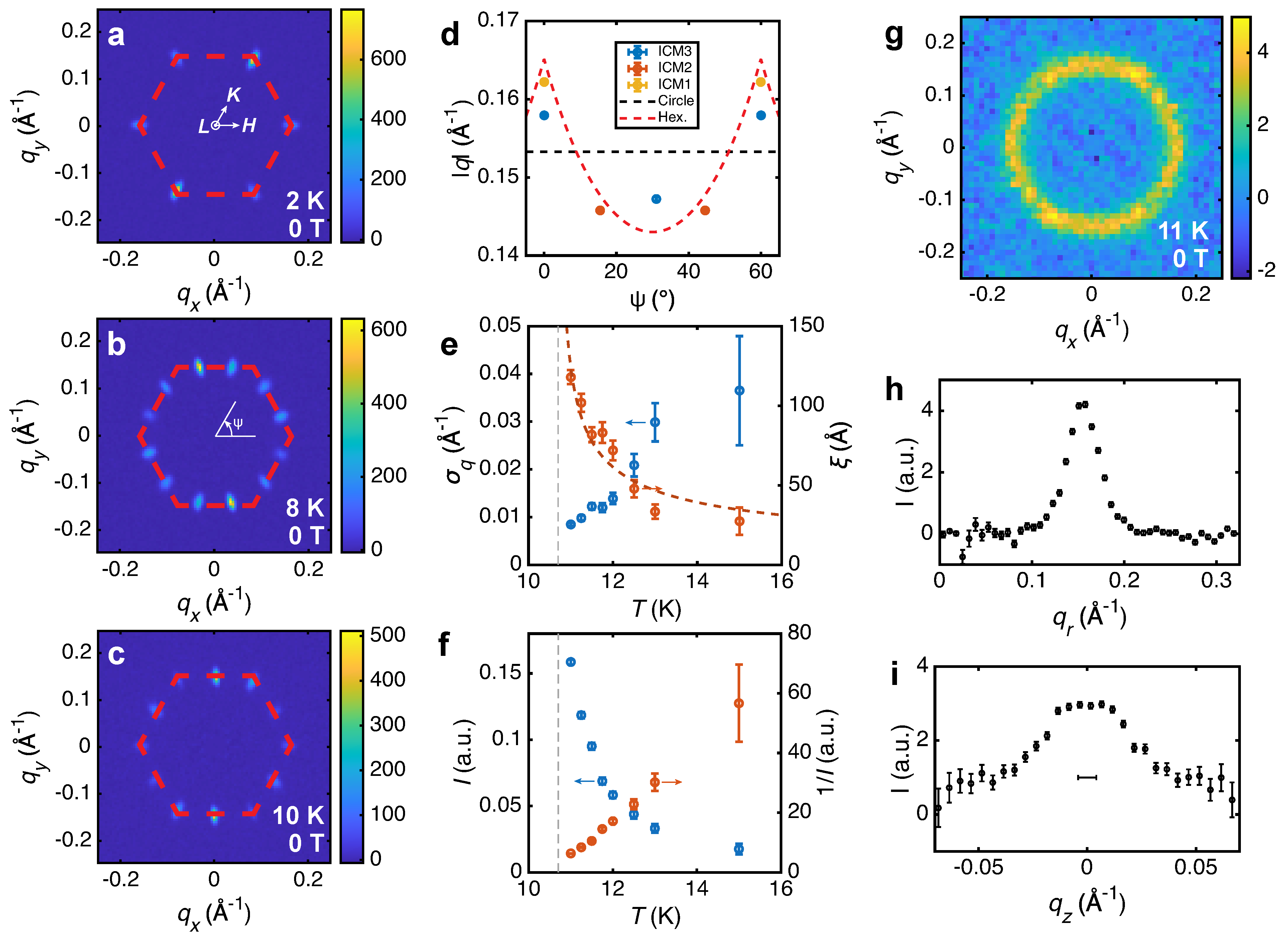}
    \caption{
    \textbf{Preferred Propagation Vectors and Critical SSL Fluctuations}
    \textbf{a-c} Temperature dependence of the ICM phases, with red hexagons emphasizing the line on which the peaks of all three phases lie, collected at zero field and 2, 8, and 10 K, respectively. 
    \textbf{d} The magnitude of the magnetic propagation vector $q$ in ICM 1-3 as a function of azimuthal angle $\psi$ (see inset of \textbf{b} for definition), where 0$^\circ$ is along the $a^*$ direction rotating counterclockwise about $c$. The expectation for a circular (hexagonal) nodal line is the dashed black (green) line.
    \textbf{e} The width $\sigma_q$, in blue, of the radially averaged diffuse scattering peak, with the instrument resolution subtracted (estimated from the width of the peaks in the ordered phase), and the correlation length $\xi=1/\sigma_q$, in orange, as a function of temperature above the ordering temperature (vertical dashed gray line), see Eq. \ref{crit} for definitions of $\xi$ and $\sigma_q$.
    \textbf{f} The integrated intensity of the radially averaged diffuse scattering as a function of temperature. \textbf{d-f} Were obtained by fitting to a Gaussian. The errorbars are the uncertainty in the fit values.
    \textbf{g} The diffuse scattering at 11 K just above the transition temperature. \textbf{h} The radial integration of the SANS intensity at 11 K. $q_r=\sqrt{q_x^2+q_y^2}$. \textbf{i} The $q_z$ dependence of the integrated SANS intensity at 11 K (the FWHM of the ground state peak is indicated with a horizontal bar as an estimate of the instrument resolution).}
    \label{qline}
\end{figure*}

First, observing the small angle neutron scattering (SANS) patterns of the three ICM phases accessible in zero field below $T_N$, a striking pattern becomes apparent. All three sets of peaks approximately lie on the same hexagon when integrated to the $q_x$-$q_y$ plane (Fig. \ref{qline}\textbf{a-c}). There are slight variations from a hexagon, see the azimuthal dependence of the magnitude of $q$ in Fig. \ref{qline}d. Further, plotting the peaks from all three phases summed together, in field and in zero field (Fig. \qlineSI\textbf{a-b}), all the peaks rest approximately on a nodal line in momentum space that modulates above and below $q_z=0$ in a manner consistent with the symmetry of the crystal's $D_{3d}$ point group (Fig. \qlineSI\textbf{c}). This suggests that there is a line in momentum space along which magnetic modulation propagation vectors are nearly degenerate.

This hypothesis is further reinforced by an examination of the diffuse scattering intensity above $T_\text{N}$. At 11 K, a broadened ring of scattering intensity is observed (Fig. \ref{qline}\textbf{g}). This ring appears in the same region as the ICM1-3 diffraction peaks and momentum space nodal line we discussed above---indicating that indeed all of these propagation vectors are similar in energy as they all fluctuate with similar intensity. Additionally, it is peaked about $q_z=0$ (Fig. \ref{qline}\textbf{i}), forming a ring of intensity in reciprocal space. This indicates that the europium moments are ferromagnetically coupled between layers, as decoupled layers would form a cylinder of diffuse intensity, and anti-ferromagnetically coupled layers would peak at $q_z\neq0$.
The width of the ring along the radial direction, see Fig. \ref{qline}h (corrected for instrument resolution, estimated from the radial width of the peaks measured in the ordered phase), $\sigma_q$, and coherence length, $\xi=1/\sigma_q$, as a function of temperature are depicted in Fig. \ref{qline}e, and the intensity of the ring is depicted in Fig. \ref{qline}f. The width fits to a critical scaling function
\begin{equation}
    \label{crit}
    \xi \propto \left(\frac{T-T_c}{T_c}\right)^{-\nu}
\end{equation}
for $\nu=0.48(07)$ (68\% confidence bounds).
Such a $\nu=1/2$ scaling has been previously discussed as the mean-field behavior in a system with a manifold of energy minima at finite momenta \cite{brazovskii1975phase, janoschek2013fluctuation}. We note that in these works, the correlation length is renormalization by critical fluctuations beyond the mean-field treatment near the transition temperature. Such deviations from mean-field behavior would be highly desirable to probe in future work.

\section{Spin Hamiltonian Modeling}
We now turn to quantitative modeling of the diffuse scattering intensity. As the nature of the spin correlations in the material is dictated by the spin interactions, a parameterization of the spin interactions can be used to model the temperature dependent diffuse SANS (see Supplementary Fig. \SANSdiffuse) and powder diffuse scattering and susceptibility (see Supplementary Fig. \powdersus) \cite{paddison2023spinteract}. Conventionally, these parameters are extracted in the ordered or field polarized state from spin modulations measured with inelastic neutron scattering, but in this case the inelastic spectrum is challenging to observe due to the large europium absorption at the neutron energies required to measure the spectrum. As a model, we consider 
\begin{equation}
\label{full_ham}
H = H_{Heis.} + H_{anis.} + H_{dip.},
\end{equation}
which contains a Heisenberg exchange term, a single ion anisotropy term, and a dipolar interaction term, respectively.
The Heisenberg exchange Hamiltonian is given by
 \begin{equation}
 H_{Heis.}=
 -\frac{1}{2} \sum_{\langle i,j\rangle}J_{ij} \mathbf{S}_{i}\cdot\mathbf{S}_{j}
 \end{equation}
 for sites indexed by $i$ and $j$, where $J_{ij}$ is the exchange coefficient and $\mathbf{S}_{i}$ are the spin on each site. The anisotropic term
 \begin{equation}
 H_{anis.}=-\Delta\sum_{i}[S_{i}^{z}]^{2}
 \end{equation}
 represents a easy-plane anisotropy in the $ab$-plane for negative anisotropic interaction $\Delta$, and an easy axis anisotropy along the $c$-axis for positive $\Delta$.
The dipole term $H_{dip}$ with interaction strength $D_{dip}$ is given as
 \begin{equation}
 H_{dip} = \frac{g^2D_{dip}|d_{NN}|^3}{2} \sum_{i} \sum_{j} \frac{\{ \bm{S}_i \cdot \bm{S}_j -3 [ \bm{S}_i \cdot \hat{\bm{r}}_{ij} ] [ \bm{S}_j \cdot \hat{\bm{r}}_{ij} ] \}}{|\bm{r}_{i,j}|^3}
 \end{equation}
where $d_{NN}$ is the nearest-neighbor distance, $g$ is the Landé $g$-factor, and $\bm{r}_{i,j}$ is the vector connecting sites $i$ and $j$. The dipole term can be estimated as $D_{dip}=\mu_0\mu_\text{B}^2g^2/(4\pi k_\text{B} d_{NN}^3)=0.00592$ K, where $\mu_0$ is the permeability of free space, $\mu_\text{B}$ is the Bohr magneton, and $k_\text{B}$ is the Boltzmann constant.
The model assumes a $S=7/2$, $L=0$ ($J=7/2$) localized spin moment with $g=2$, in accordance with the Eu$^{2+}$ half-filled $f$ shell. The expectation value of the magnitude of the moment is $g\mu_\text{B}\sqrt{J(J+1)}=7.94\mu_\text{B}$, which matches well with the Curie-Weiss effective moment of $8.09\mu_\text{B}$ extracted from the powder susceptibility. 
By fitting the spin interaction parameters until the model matches the observed diffuse scattering, susceptibility, and saturation fields, we can obtain an experimental model of the spin interactions as has been performed in related compounds \cite{paddison2022magnetic}.

\begin{figure*}[ht]
\includegraphics[width=1.0\textwidth]{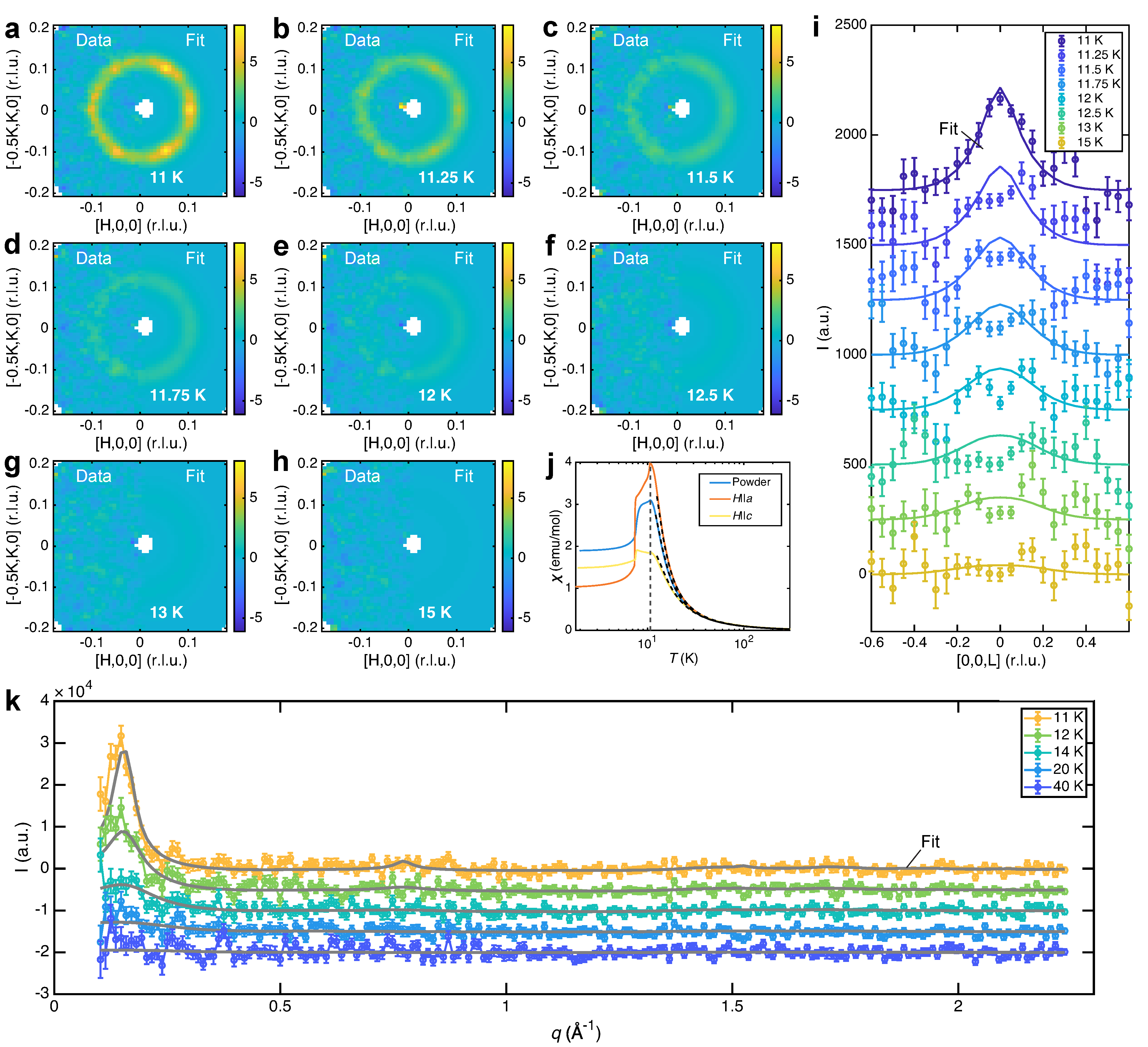}
    \caption{
    \textbf{Model of Diffuse Scattering and Susceptibility}
    \textbf{a-h} Fit of diffuse single crystal SANS intensity to Heisenberg model as a function of temperature (see text, table \ref{tab:Jr}, and Supplemental Information for details). The data is shown on the left half of each figure, while the model is depicted on the right half of each figure. \textbf{i} Model of the $L$-dependence of the single crystal SANS scattering intensity. The data is fit in a 3D cube, and plotted here as a integral over the $H$ and $K$ directions.
    \textbf{j} Model fit of susceptibility data. \textbf{k} Model fit of the diffuse powder diffraction data. The SANS data is binned, smoothed, inversion symmetrized, and a high temperature (20 K) background is subtracted. The powder data has a high temperature (100 K) background subtracted. Single crystal susceptibility data is reproduced from \cite{kurumaji2025electronic}.}
    \label{Jr}
\end{figure*}

A simultaneous fit of the temperature dependent SANS (Fig. \ref{Jr}\textbf{a-i}), powder and single crystal susceptibility (Fig. \ref{Jr}\textbf{j}), and temperature dependent neutron powder diffraction data (Fig. \ref{Jr}\textbf{k}) was conducted to optimize the reaction field model \cite{paddison2023spinteract, logan1995onsager}. The first out-of-plane Heisenberg interaction ($J_3$) was included, along with the 8 nearest neighbor in-plane Heisenberg interactions, the anisotropy term, and a fixed dipole term. The optimized parameters (Table \ref{tab:Jr}) generate a good fit to the data. Adding more in-plane Heisenberg parameters does not significantly improve the quality of the model (see Supplementary Fig. \chisqvsparams). The model also accurately predicts the experimental Néel temperature of 10.7 K. 

\begin{table}[t]
\caption[Diffuse Scattering Model Fit Parameters]{Fit parameters for the model of diffuse scattering and susceptibility, in K. The exchange energy can be converted from K to meV by multiplying by $k_{B}$=0.08617 meV/K. The errors presented in this table are the statistical errors from the global fit.}\label{tab:Jr}%
\centering{%
\begin{tabular*}{0.25\textwidth}{@{\hspace*{1.5em}}@{\extracolsep{\fill}}cc@{\hspace*{1.5em}}}
\\[-0.5em]
\toprule
Term & Energy (K) \\
 \midrule
$D_{dip.}$ & 0.00592 \\
$\Delta$ & -0.2262(8) \\
$J_{1}$ & 0.3423(7) \\
$J_{2}$ & -0.0200(6) \\
$J_{4}$ & -0.0783(5) \\
$J_{7}$ & 0.0671(3) \\
$J_{10}$ & 0.1134(5) \\
$J_{13}$ & 0.0505(4) \\
$J_{16}$ & -0.1776(2) \\
$J_{21}$ & 0.0497(3) \\
$J_{3}$ & 0.1553(3) \\
\bottomrule
\end{tabular*}
}%
\end{table}

The nearest neighbor interaction in the europium layer is ferromagnetic, while later terms oscillate. The frustration of these various interactions pushes the system to prefer incommensurate order.
While the oscillatory nature of $J(r)$ is generally reminiscent of the familiar RKKY interaction picture, the oscillation frequency is much too fast for the experimentally observed $q=2k_\text{F}$ (order $\sim$1 nm instead of $\sim$4 nm, see Supplementary Fig. {\chisqvsparams}g) and does not follow the functional form. However, to accurately implement the expected RKKY model would require including $J_i$ terms to distances greatly exceeding $2\pi/q\approx40$ \AA\ at a minimum or performing the model calculations in momentum space as has been discussed \cite{fuchizaki1994towards}. The truncated model here may be considered an effective one which approximates the true $J(r)$ at low energy and to limited resolution in the momentum space vector $Q$.

\section{Momentum Space Energy Landscape}
\begin{figure}[htb]
\includegraphics[width=1.0\columnwidth]{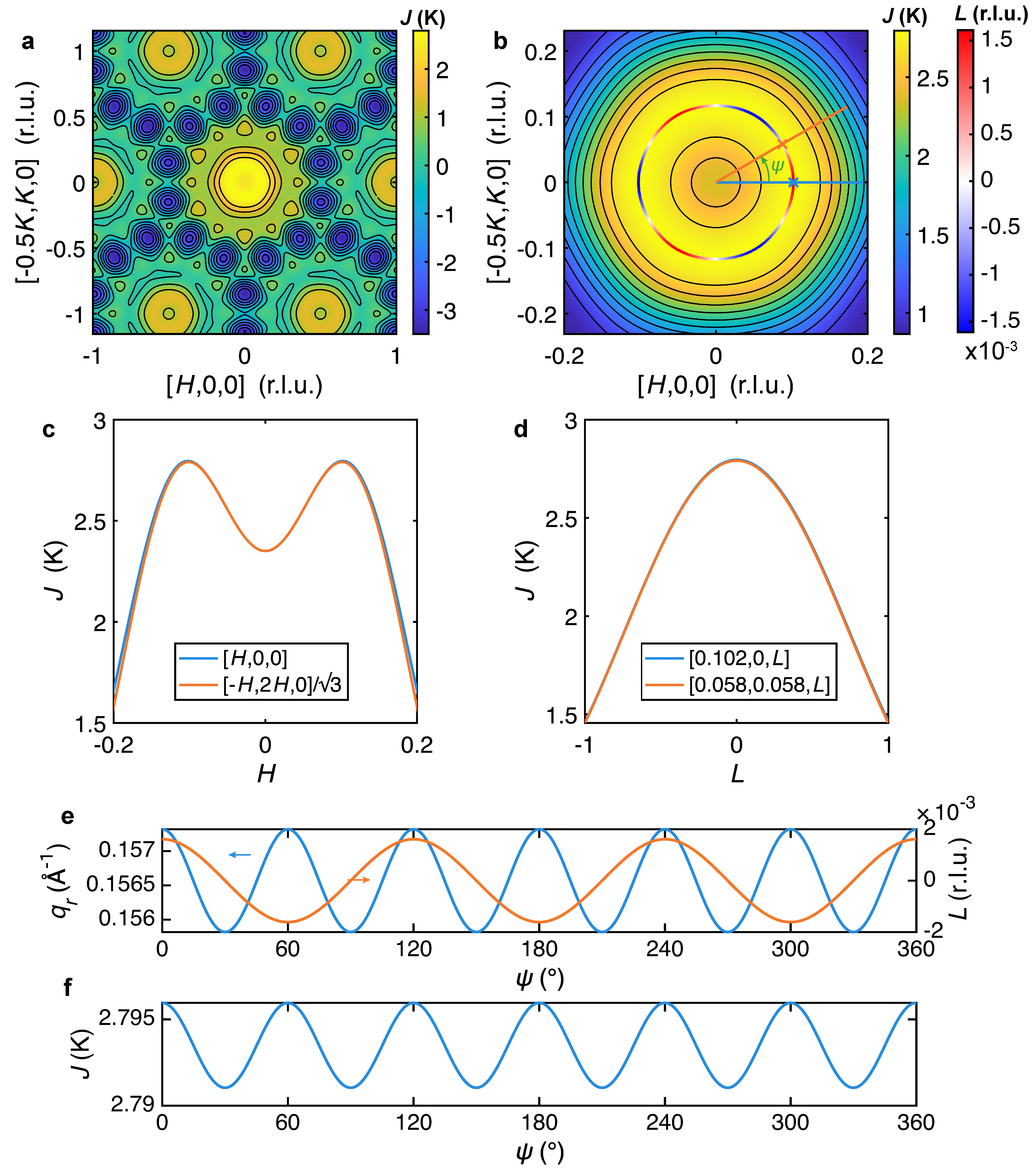}
    \caption{
    \textbf{Momentum Space Pairing $J(\bm{Q})$}
    The maximum eigenvalue of the interaction matrix in reciprocal space, $J(\bm{Q})$ plotted \textbf{a-b} in the $HK0$ plane, \textbf{c-d} along the $(H,0,0)$ and $(-H/2,H,0)$ directions, and \textbf{e} along (0.102, 0, $L$) in blue and (0.058, 0.058, $L$) in orange. The line cuts plotted in \textbf{c} are indicated in \textbf{b} by lines, and the intersections of the line cuts plotted in \textbf{d} are indicated with crosses in \textbf{b}. In \textbf{b}, the location of the maximum as a function of in-plane angle is indicated with a red-white-blue line, with the color of the line indicating the component along $L$.
    \textbf{e} The radial in-plane distance of the maximum (along the red line in \textbf{b}) of $J(\bm{Q})$ from the origin in \AA$^{-1}$ as a function of in-plane angle $\theta$, where $\theta=0$ corresponds to the $a^*$ direction.
    \textbf{f} The value of $J(\bm{Q})$ at the maximum (along the red line in \textbf{b}) as a function of in-plane angle $\theta$.
    See supplementary information for additional cuts of $J(\bm{Q})$.}
    \label{Jk}
\end{figure}

In order to understand the energy landscape and its connection to the order in this compound better, we calculate the maximum eigenvalue of the interaction matrix in momentum space, $J(\bm{Q})$.
Plotting the momentum-space $J(\bm{Q})$ (Fig. \ref{Jk}a-b), we observe a broad flat peak in strength near the origin, which is slightly peaked (Fig. \ref{Jk}c) along a circle that is consistent with the location of the diffuse scattering intensity. The strength of $J(\bm{Q})$ also decays away from the $L=0$ plane (Fig. \ref{Jk}d) as is observed with the diffuse scattering (see supplementary section SIV for additional cuts of $J(\bm{Q})$). Further, the peak in $J(\bm{Q})$ occurs at $\bm{Q}_{max}=(0.102,0,0.002)$ which is in good agreement with the location of the stronger of the two propagation vectors at $(0.103,0,0.020)$ observed in ICM3 in SANS. 
Just above the ordering temperature, $\chi(\bm{Q})$ for this model yields a well-defined ring centered on the $q_z=0$ plane which matches well with the observed diffuse intensity (see SANS intensity in Fig. \ref{Jr}g-i). This ring is also consistent with the fact that the system can order along many different ordering vectors with $|\bm{q}|\sim0.16$ \AA$^{-1}$, depending sensitively on the exact temperature and applied field. The model also qualitatively reproduces the azimuthal dependence of the in-plane and out-of-plane components of the propagation vectors (see Fig. \ref{Jk}e).

In the reaction field theory, the system will order along the first $\bm{Q}$ that causes $\chi(\bm{Q})$ to diverge as temperature is lowered (which will occur at $\bm{Q}_{max}$). Neglecting higher order interactions, the ground state of the system will be to order with the propagation vector at the maximum of $J(\bm{Q})$ which in this model slightly prefers (0.102, 0, 0.002) over (0.058, 0.058, 0) (compare the blue and orange cross in Fig. \ref{Jk}b, respectively) --- the exchange energy is 2.9225 K vs 2.9175 K, only a 0.17\% difference (see azimuthal dependence of $J$ in Fig. \ref{Jk}f). This is in remarkably good agreement with the observed ordering in ICM1 at 2 K of $q_{ICM1}=$(0.105, 0, 0.055) and ICM3 at 10 K of $q_{1,ICM3}=$(0.103, 0, 0.02) and $q_{2,ICM3}=$(0.056, 0.056, 0) \cite{kurumaji2025electronic}. Higher order interactions like a four-spin interaction may favor multi-$q$ order \cite{hayami2017effective, hayami2024stabilization, forgan1989observation, khanh2020nanometric,Takag:2022}, anisotropy and the saturated moment condition may dictate which type of spin modulation is favored, and the Zeeman energy from applied magnetic fields will alter the energy landscape further.

\section{Monte Carlo Simulations}
\begin{figure}[htb]
\includegraphics[width=1.0\columnwidth]{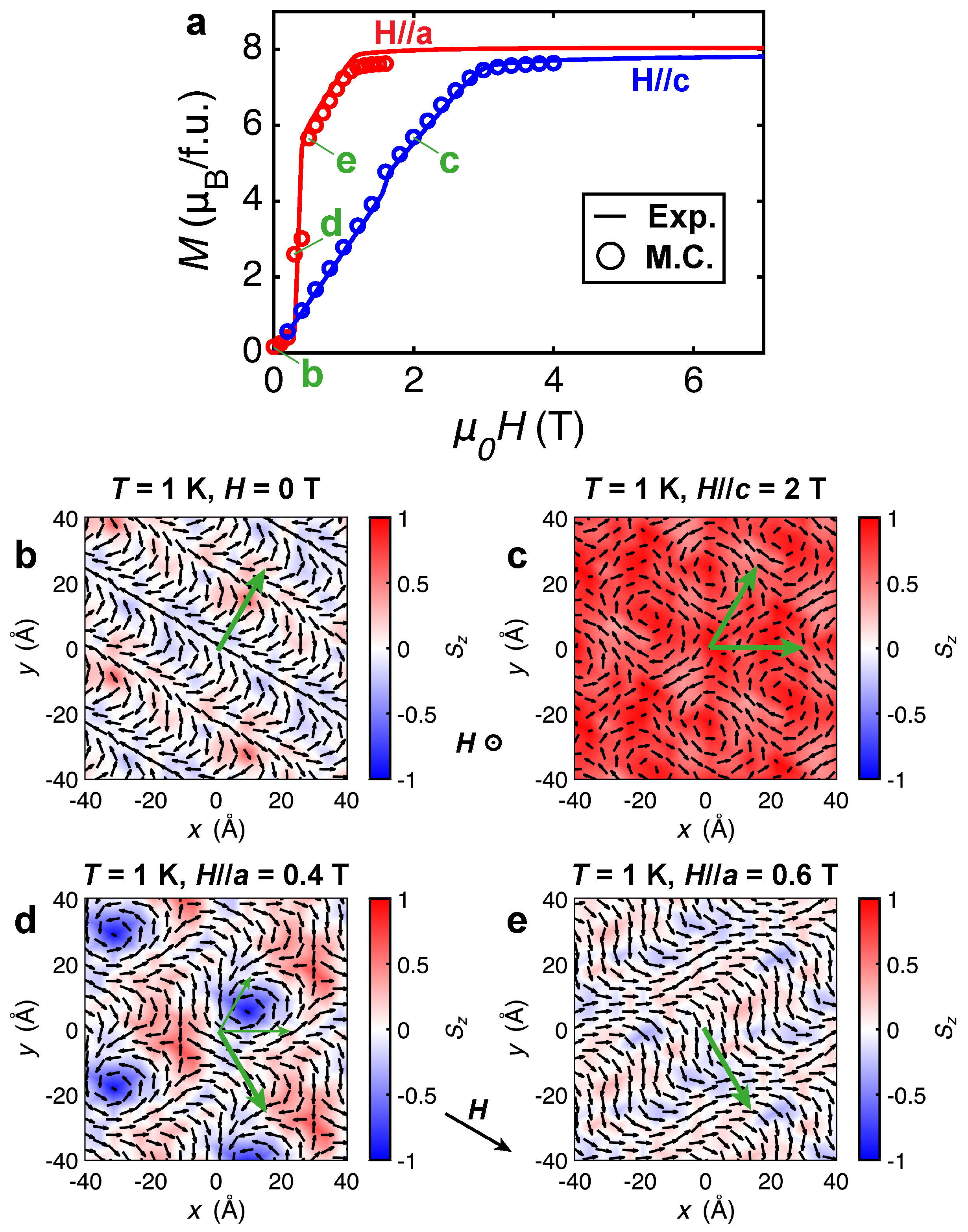}
    \caption{
    \textbf{Monte Carlo Simulation}
    \textbf{a} The MC simulated (circles) and experimental (lines) magnetization as a function of applied field along the $a$ (red) and $c$ (blue) directions at 2 K. The temperature-dependent phase diagrams are compared in supplementary Fig. \MCPD. \textbf{b-e} The MC simulated magnetic textures at 1 K for various applied fields (indicated by green letters in \textbf{a}). The field directions are depicted with black arrows, and the magnetic propagation vectors are depicted with green arrows. One $ab$-plane layer of magnetic spins is depicted, with the in-plane magnetization shown with black arrows, and the out-of-plane magnetization depicted with the colorscale. The spin size is normalized to one. \textbf{b} is simulated in zero field, \textbf{c} is simulated with 2 T of field along $c$, and \textbf{d,e} are simulated with 0.4 T and 0.8 T of field along $a$, respectively. These represent the four states observed in our MC modeling. The diffraction patterns corresponding to \textbf{b-e} are depicted in supplementary Fig. \MCdiff. Each 10x10 simulation was periodically tiled for visual clarity. Single crystal magnetization data is reproduced from \cite{kurumaji2025electronic}.}
    \label{MC_result}
\end{figure}

The utility of this model can be further tested by comparing MC modeling of the ordered behavior to the experimental data. As the model was fit only with data in the paramagnetic state and the saturation fields, simulating of the ordered states tests the model’s validity beyond the fitted domain. Our simulation rather accurately reproduces the magnetization as a function of applied field as measured at 2 K (Fig. \ref{MC_result}a). Both the saturation field and observed metamagnetic transitions at $\bm{H}||\bm{a}\sim$0.35 T and $\bm{H}||\bm{c}\sim$1.5 T are reproduced. The model actually predicts two metamagnetic transitions at $\bm{H}||\bm{a}=$0.25 T and $\bm{H}||\bm{c}=$0.45 T instead of one, which is only experimentally observed at slightly higher temperatures \cite{green2025robust, neves2025inplane}. The experimentally observed intermediate phase does extend down close to low temperature, so it is likely very close in energy in the real material at 2 K at this field. The ICM2 to ICM3 phase transition is not observed at $\bm{H}||\bm{c}\sim$2.5 T, but in previous work it was shown that a four-spin interaction is necessary to explain the coexistence of both ICM2 and ICM3 in the same phase diagram, which is not included in this model \cite{neves2025polarized}.

Turning now to the real-space simulated spin textures, the MC model correctly predicts a zero field ground state which is a in-plane single-$q$ cycloid (Fig. \ref{MC_result}b). The simulation also accurately predicts the presence of a double-$q$ vortex-antivortex lattice for out-of-plane field, and the vortex cores become saturated out of plane (Fig. \ref{MC_result}c). This simulation predicts a lattice with propagation vectors which are 60$^\circ$ apart, while in the real material they are closer to (ICM2) and exactly (ICM3) 90$^\circ$ apart. This discrepancy may be due to finite size effects of the limited simulation size, or some other higher order interaction not considered here. Regardless, it is remarkable that the simulation is accurate in its prediction of ICM2/ICM3.

For applied field along $a$, the simulation predicts first a triple-$q$ phase where the dominant propagation vector is close to the applied field direction. This phase is indeed experimentally observed with SANS as ICM2c \cite{neves2025inplane} though only with field along $a^*$, though the real-space texture has not been experimentally confirmed. Intriguingly, this texture is predicted here to have a finite scalar spin chirality, containing merons and antimerons in 1D lines. Experimental confirmation of this is therefore highly interesting. The experimentally observed ICM2a-b phase is not present in this model. This may be due similarly to the lack of a four-spin interaction term, or some other small correction. Finally, the model indeed predicts a single-$q$ texture for higher in-plane field which was also predicted by previous modeling \cite{neves2025inplane}. Additional plots of the simulated phase diagram and SANS diffraction patterns are presented in supplementary section SV.

This model is qualitatively similar to the model previously developed \cite{neves2025polarized}. Both can predict the phases with a small easy-plane anisotropy (the $zz$-term of the interaction matrix is about $90\%$ the magnitude of the $xx$- and $yy$-terms), and both contain some small cross $xy$ terms in the interaction matrix. However, while that model was hand tuned and only included interaction matrices at the primary propagation vectors, this model is experimentally refined and simulates the interaction matrix everywhere. This model also accurately includes the $\bm{Q}=0$ interactions, which therefore enables accurate simulation of the anisotropic magnetization.

It is remarkable that this model accurately predicts the magnetization and four out of six of the observed magnetic phases in this complex material. This represents, to our knowledge, the first time such an accurate model of such a complex system has been obtained without inelastic scattering data. Such an approach is thus promising moving forward, especially with small sample sizes or absorbing samples where spectroscopy may be challenging. Future extensions may therefore be of interest that contain multi-spin interactions, or more direct connections with the mediating Fermi surface.

This represents a general design principle for the engineering of spin moir\'e superlattice materials and metallic spiral spin liquids. A clean 2D lattice of local moments coupled with a quasi-2D conduction band (where ideally the conduction electron orbitals overlap strongly in real space with the local moment orbitals to enhance the coupling), where one band is cylindrical with the $k_F=q/2$ of choice, will produce a landscape of proximate multi-$q$ spin textures. The magnetic ordering of these textures in turn alters the electronic structure, producing strong changes in magneto-electrical properties. Additionally, the coupled real and momentum space topology generated by these phases may lead to a variety of quantum and anomalous Hall effects of potential technological relevance.

\section{Methods}

\subsection{Synthesis}
Single crystals were synthesized via a self-flux method described in \cite{kurumaji2025electronic}. Powder samples used for neutron powder diffraction were synthesized as described in \cite{gerke2013magnetic}.

\subsection{Small Angle Neutron Scattering}
Diffuse small angle neutron scattering measurements were performed with GP-SANS at the High Flux Isotope Reactor at Oak Ridge National Laboratory using 4.0 \AA\ neutrons. Preliminary diffuse measurements and the data in ICM1-3 shown in Supplementary Fig. \qlineSI a-b were performed with SANS-I at the Swiss Spallation Neutron Source at the Paul Scherrer Institut using 3.1 \AA. The sample was rocked about the vertical and/or horizontal axes to map a 3D volume of reciprocal space. SANS measurements were analyzed in GRASP \cite{dewhurst2023graphical} and in the GRIP module \cite{neves2024grasp}. Every measurement has a 20 K background subtracted.

\subsection{Neutron Powder Diffraction}
Powder diffraction was performed with the HB-2A powder diffractometer at the High Flux Isotope Reactor at Oak Ridge National Laboratory using 2.41 \AA\ neutrons. The instrument was used in an open-open-12' collimation setting (only the pre-sample collimator was used). The $\sim$1 g sample was placed in a pressed aluminum foil sachet and sealed into an aluminum can filled with helium exchange gas. This was done to produce an annular sample geometry, which was chosen to reduce the effect of absorption from Eu. Powder refinements were performed with Mag2Pol \cite{qureshi2019mag2pol}.

\subsection{Magnetization Measurements}
Bulk magnetic susceptibility and magnetization measurements were performed in a Quantum Design Magnetic Property Magnetization System 3.


\subsection{Diffuse Scattering Modeling}
The diffuse scattering modeling was performed using Spinteract \cite{paddison2023spinteract}. The fit was performed simultaneously on the SANS, powder diffraction, powder and single crystal susceptibility data, and the minimum ($\bm{H}||\bm{a}$) and maximum ($\bm{H}||\bm{c}$) saturation fields. The model contained the six nearest neighbor in-plane and nearest neighbor out-of-plane Heisenberg interactions (see Fig. \ref{intro}a), an anisotropic term, and a dipole term (see Eq. \ref{full_ham}). The inclusion of further-neighbor interactions in-plane did not significantly alter the fit quality or conclusions (see Supplementary Fig. \chisqvsparams). Additional details regarding the reaction field theory may be found in supplemental section SV.

\subsection{Monte Carlo Modeling}
The Monte Carlo models were performed on a 10x10x2 grid containing 600 atoms using a custom Monte Carlo code. The spin Hamiltonian is given by Eq. \ref{full_ham}, and the interaction parameters are specified in Table \ref{tab:Jr}. Spins were modeled as classical vectors of length $\sqrt{S(S+1)}$ with $S=7/2$. Simulations were initialized at $T=$20 K and cooled in 1 K steps to $T=$1 K in a fixed applied magnetic field parallel to $\bm{a}$ or $\bm{c}$. At each temperature, the number of proposed Monte Carlo steps, $N_d$, needed to de-correlate the spin system was estimated. Simulations were run for $10N_d$ steps for equilibration followed by $1000N_d$ steps for measurement. A single proposed step comprised an over-relaxation move of a single spin followed by a rotation of this spin. Proposed moves were accepted or rejected according to the Metropolis algorithm. The long-range dipolar interaction was implemented using Ewald summation, which restricts the simulations to relatively small sizes. The neutron diffraction patterns were simulated with Scatty \cite{paddison2019ultrafast}. In the plotted textures and diffraction images, the unit cell was tiled periodically to improve the clarity of the visualization.


\subsection{Acknowledgments}
We appreciate fruitful discussions with C. John, S. Moody, J. Lass, N. P. Butch, A. Minelli, S. Hayami, M. Aronson, I. Martin, A. Rosch, I. Mazin, and C. Broholm.
This work was funded, in part, by the Gordon and Betty Moore Foundation EPiQS Initiative, grant no. GBMF9070 to J.G.C. (instrumentation development); the US Department of Energy (DOE) Office of Science, Basic Energy Sciences, under award no. DE-SC0022028 (material development); the Office of Naval Research (ONR) under award no. N00014-21-1-2591 (advanced characterization); and the Air Force Office of Scientific Research (AFOSR) under award no. FA9550-22-1-0432 (magnetic structure analysis).
A portion of this research used resources at the High Flux Isotope Reactor, a DOE Office of Science User Facility operated by the Oak Ridge National Laboratory. The beam time was allocated to GP-SANS on proposal number IPTS-32248.1, and to HB-2A on proposal number IPTS-33869.1. J.S.W. acknowledges financial support from the Laboratory for Neutron Scattering and Imaging at PSI for an extended research visit of P.M.N. This work is based partly on experiments performed at the Swiss Spallation neutron source SINQ, Paul Scherrer Institute, Villigen, Switzerland.

\subsection{Contributions}
Small angle neutron scattering was performed by P.M.N. with support from L.M.D.-S., and J.S.W.. Powder neutron diffraction was performed by P.M.N.and C.I.J.I. with support from J.A.M.P.. Materials were synthesized and characterized by T. K. and C.I.J.I.. S.F. performed DFT calculations. Lindhard calculations were performed by P.M.N.. Diffuse scattering modeling was performed by P.M.N. and J.A.M.P.. Monte Carlo modeling was performed by J.A.M.P. and P.M.N.. Spectroscopic measurements were performed by P.M.N. and D.G.M.. All authors contributed to writing the manuscript. J.G.C. coordinated the project.




%

\end{document}